\documentclass{elsart}

 \usepackage{graphics}
 \usepackage{graphicx}

\usepackage{amssymb}

\begin{document}

\begin{frontmatter}



\title{PQCD approach to parton propagation \\ in matter}


\author{Ivan Vitev}

\ead{ivitev@lanl.gov}

\address{ Los Alamos National Laboratory, 
Theoretical Division and Physics Division \\ Los Alamos, NM 87545, USA }

\begin{abstract}

We review recent theoretical developments in understanding 
the many-body perturbative QCD dynamics of strongly-interacting 
hard probes in dense nuclear matter. The relation   
between initial- and final-state parton scattering in 
nuclei and the quark-gluon plasma and experimental measurements
of the Cronin effect, nuclear shadowing, jet quenching, 
modified di-hadron correlations and forward rapidity particle 
suppression in p+A and A+A reactions is clarified. Our 
approach emphasizes the process dependence of nuclear effects 
and outlines the techniques for their dynamical calculation and 
incorporation in simulations that build upon collinear 
factorization in QCD.

\end{abstract}


\end{frontmatter}

\section{Introduction}
\label{intro}

In high energy p+p collisions jet and particle production 
is well described in the framework of the collinear PQCD
factorization approach. However, when a partonic process is 
embedded in nuclear matter, the cross sections are modified 
by the  soft scatterings that precede and/or follow the hard 
interaction~\cite{Vitev:2006bi}. One can gain insight 
in the various aspects of perturbative many-body QCD 
dynamics by following the history of a parton from 
early times, $t\rightarrow -\infty$,
in the parton distribution function to its asymptotic 
state, $t \rightarrow \infty$, in the wave function of 
the observed hadron. In our approach, the strength of 
jet interactions is controlled by microscopic and 
non-perturbative parameters  characterizing the medium, 
such as temperature $T$, energy density $\epsilon$, the
scale of high twist corrections $\xi^2$, typical momentum 
transfer squared $\mu^2$, and mean free path $\lambda$.  
Results are summarized~\cite{Vitev:2006bi} in 
Table~\ref{class-1}.

\begin{table}[!t]
{ \begin{tabular}{||l|l||}
\hline
{\bf Type of  scattering}  \ \ \ \  &
{\bf $p_T$ dependence of the nuclear effect} \\ [.5ex]
\hline
{\it Elastic (incoherent)} $\qquad$   & 
I1P:  Cronin effect; I2P: di-jet acoplanarity  \\ [.5ex] %
{\it Inelastic (final-state)} & I1P: quenching at all $p_T$; 
 I2P:  quenching/enhancement  \\ 
 & at high/low $p_T$ \\ [.5ex]
{\it Inelastic (initial-state)} & 
I1P:  suppression at all $p_T$;  I2P: same as I1P \\[.5ex]  
{\it Coherent ($t-$, $u-$channel)}  & I1P:
suppression at low $p_T$; I2P: same as I1P  \\[.5ex]
{\it Coherent ($s-$channel)}  & I1P: 
enhancement at low $p_T$; 
I2P: same as I1P \\
\hline
\end{tabular}
\vspace*{.2cm}
\caption{ Effect of elastic, inelastic and coherent multiple 
scattering on  the transverse momentum dependence of single (I1P) 
and  double inclusive (I2P) hadron production in the 
perturbative regime. } }
\label{class-1}
\end{table}

\section{Cronin effect and dynamical shadowing}
\label{collisional}

\begin{figure}[!t]
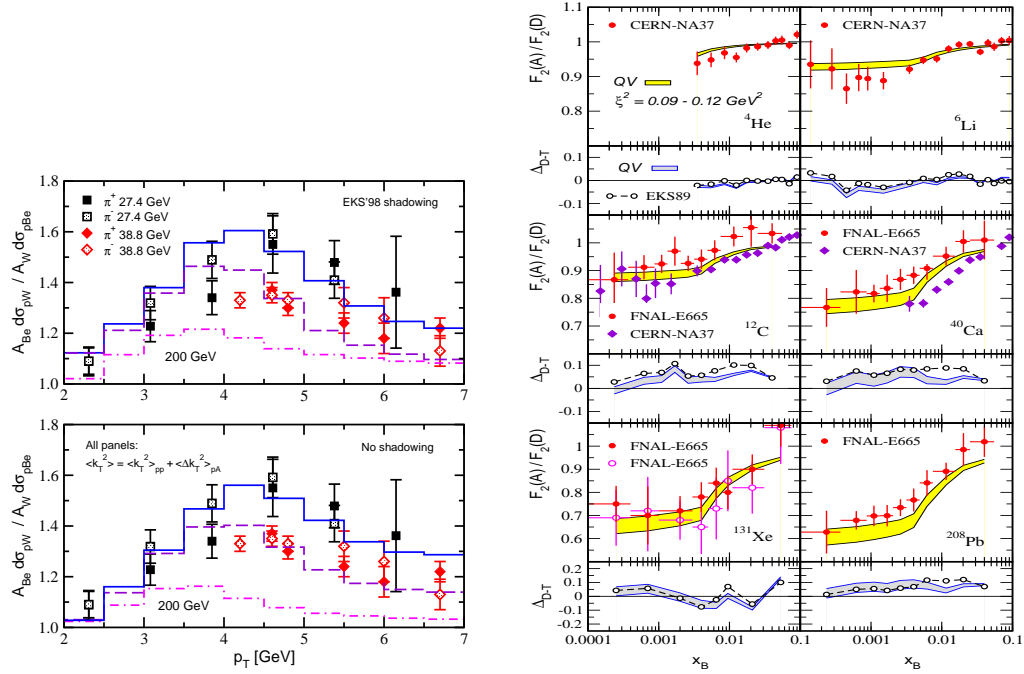

\includegraphics[width=2.4in,height=2.6in,angle=0]{cron.SHnoSH.eps}
\hspace*{0.5cm}
\includegraphics[width=2.6in,height=3.5in,angle=0]{Shadow.eps}
\caption{ Left panel: the ratio of $A$-scaled  p+W/p+Be data 
on $\pi^+,\pi^-$ production at  $\sqrt{s}=27.4, 38.8$~GeV. 
Right panel: all-twist resummed $F_2(A)/F_2(D)$ calculation.
The band corresponds to the choice $\xi^2 = 0.09 - 0.12$~GeV$^2$. 
}
\label{elast}
\end{figure}

We first consider partonic $2 \rightarrow 2$ processes, which 
we call elastic as opposed to inelastic $2 \rightarrow 2+n$
processes associated with gluon bremsstrahlung. Multiple 
interactions of the incoming quarks and gluons lead to 
transverse momentum diffusion~\cite{Vitev:2003xu}. 
For initial parton flux  ${d^2N^{i}({\bf k})} = 
\delta^2({\bf k} ) d^2 {\bf k}$ the 
distribution of partons at the hard collision vertex reads:  
\begin{eqnarray}
\frac{d^2N^{f}({\bf k})}{ d^2 {\bf k} }
& = &  \sum_{n=0}^{\infty} \frac{\chi^n}{n!} 
\int  \prod_{i=1}^n   d^2 {\bf q}_{i} 
 \left[  \frac{1}{\sigma_{el}} 
\frac{d\sigma_{el} }{d^2{\bf q}_{i} }
 \left(   e^{-{\bf q}_{i} \cdot 
\stackrel{\rightarrow}{\nabla}_{{\bf k} }}  
 - 1  \right) \, \right] 
\frac{d^2N^{i}({\bf k})}{ d^2 {\bf k} }
  \nonumber \\   
\; && \approx 
 \frac{1}{2\pi} \frac{e^{-\frac{{\bf k}^2 }{2\,\chi \,\mu^2 \xi}}}
 {\chi\, \mu^2\, \xi } \;, \qquad \chi = \frac{L}{\lambda}, \;\;
\xi = \xi({\bf k}^2)\;. 
\label{ropit}
\end{eqnarray} 
Such broadening can be implemented in the PQCD formalism 
as follows:
\begin{equation}
\langle {\bf k}^2_i \rangle =  \langle {\bf k}^2_{i\, {\rm vac}}
\rangle   + 2 \chi_i  \mu^2 \, \xi, \; \quad i = 1,2 \;, 
\end{equation}
and leads to $\sqrt{s}$-dependent enhancement of the hadron 
cross sections at intermediate $p_T \sim\;$few GeV. The 
left panel of Fig.~\ref{elast} shows theoretical calculations
at $y=0$ compared to  p+W/p+Be data and  
predictions for  d+Au collisions at RHIC. We used 
 $2 \mu^2/\lambda_q  \approx 0.12 \; {\rm GeV}^2/{\rm fm}$,   
$2 \mu^2/\lambda_g  \approx 0.28  \; {\rm GeV}^2/{\rm fm} $.
For a survey of models of the Cronin effect see~\cite{Accardi:2002ik}.

A distinctly different effect arises from coherent final-state 
interactions in large nuclei~\cite{Qiu:2003vd}. 
When the light cone momentum fraction 
$ x_B < 0.1 $ the longitudinal extent of the exchange  
virtual meson $l_c = 1 / 2 x_B m_N $ exceeds the nucleon size 
$r_0$ and the struck parton scatters coherently on several 
nucleons. The high twist scale per nucleon is  defined as:
\begin{eqnarray}
\xi^2  & = &  \frac{3 \pi \alpha_s(Q^2)}{8\, r_0^2} 
\langle p| \, \hat{F}^2 (\lambda_i) \,| p \rangle  
=  \frac{3 \pi \alpha_s(Q^2)}{8\, r_0^2} \lim_{x\rightarrow 0}
\frac{1}{2}x G(x)\;.  
\label{xi2}
\end{eqnarray}
Nuclear shadowing can be  cleanly  studied 
only in deep inelastic scattering (DIS). 
Resummation of all nuclear-enhanced power corrections  
to the inclusive cross sections generates dynamical mass 
for the final-state parton $m^2_{\rm dyn} = \xi^2 A^{1/3} $.
On the example of the transverse structure function
we have: 
\begin{eqnarray} 
\hspace*{-.7cm}
F_T^A(x,Q^2) \! & \approx &  A \, F_T^{\rm (LT)}\left( x + 
\frac{x \, \xi^2 ( A^{1/3}-1) }{Q^2},    Q^2 \right)  
= F_T^{\rm (LT)}\left( x + 
\frac{x m^2_{\rm dyn} }{Q^2},    Q^2 \right)  \; . \quad
\label{FTres}   
\end{eqnarray} 
The left panel of Fig.~\ref{elast} shows good description of 
the world's data on DIS on nuclei~\cite{Qiu:2003vd} with
$\xi^2 = 0.09 - 0.12$~GeV$^2$ and $Q^2 \geq m_N^2 $.
Similar results were derived for  p+A reactions~\cite{Qiu:2004da}
and our calculations have been generalized to include the case
of heavy quarks~\cite{Vitev:2006bi}. 
For a survey of models of nuclear shadowing see~\cite{Accardi:2002ik}.

Initial-state multiple scattering in the coherent regime, in 
contrast to final-state, leads to enhancement of the cross sections 
and  Cronin effect~\cite{Vitev:2006bi}. Such distinctly different 
behavior points to the process dependence
of nuclear effects and the need for their dynamical evaluation.

\section{Initial- and final-state energy loss}
\label{radiative}

The acceleration of the incoming and outgoing partons 
by the soft interactions in the dense nuclear matter results in 
medium-induced gluon bremsstrahlung~\cite{Gyulassy:2003mc}.
Given its quantitative importance for the description of the 
large jet attenuation in a quark-gluon plasma (QGP) of initial 
temperature $T_0 \sim 400$~MeV and energy density $\epsilon_0 
\sim 20$~GeV/fm$^3$, recent perturbative calculations have 
been focused on final-state radiative energy loss. The goal of 
such studies is to describe the large five-fold suppression 
of high-$p_T$ hadron production in central Au+Au 
reactions~\cite{Shimomura:2005en,Dunlop:2005xe}.

We here rely on the Gyulassy-Levai-Vitev (GLV) 
approach for describing the multiple interactions of the propagating 
jet+gluon system~\cite{Gyulassy:2000er,Gyulassy:2000fs} since it 
was specifically developed to describe the QCD dynamics of heavy
ion (p+A and A+A) reactions. It is well equipped to account for 
some of the most important characteristics of nuclear collisions: 
$L/\lambda_g \leq {\rm few} \ll \infty$ and $E_{\rm jet} \ll 
\infty$.  To first order in the correlation between multiple 
scattering centers  the  differential bremsstrahlung spectrum 
is given by~\cite{Gyulassy:2000er}:  
\begin{eqnarray}
\omega\frac{dN_g}{d\omega\, d^2 {\bf k}}  &=&
\frac{ C_R \alpha_s}{  \pi^2 } 
\int_{z_0}^L \frac{d \Delta z}{\lambda_g(z)}  
\int  d^2 {\bf q} \, 
\frac{\mu^2(z)}{\pi({\bf q}^2 + \mu^2(z) )^2} \times  \nonumber \\ 
&& \hspace*{-1cm} \times 
\frac{2 ({\bf k} - {\bf q}_{\rm dir.}) 
\cdot{\bf q} }{ ({\bf k} - {\bf q}_{\rm dir.} )^2 
({\bf k} - {\bf q}_{\rm dir.} - {\bf q})^2 }
\left[ 1 - \cos \left( 
\frac{ ({\bf k} - {\bf q}_{\rm dir.} -  
{\bf q})^2 \Delta z}{2\omega} \right) \right]
\;.
\label{1storderSh}
\end{eqnarray}
Here ${\bf q}_{\rm dir.}$ is possible $\neq 0$ directed 
momentum transfer, for example transverse flow. Clearly, this effect amounts
to a change of the reference frame for the radiative spectrum
but neither distorts it nor changes the amount of lost 
energy $\Delta E$.

\begin{figure}[!t]
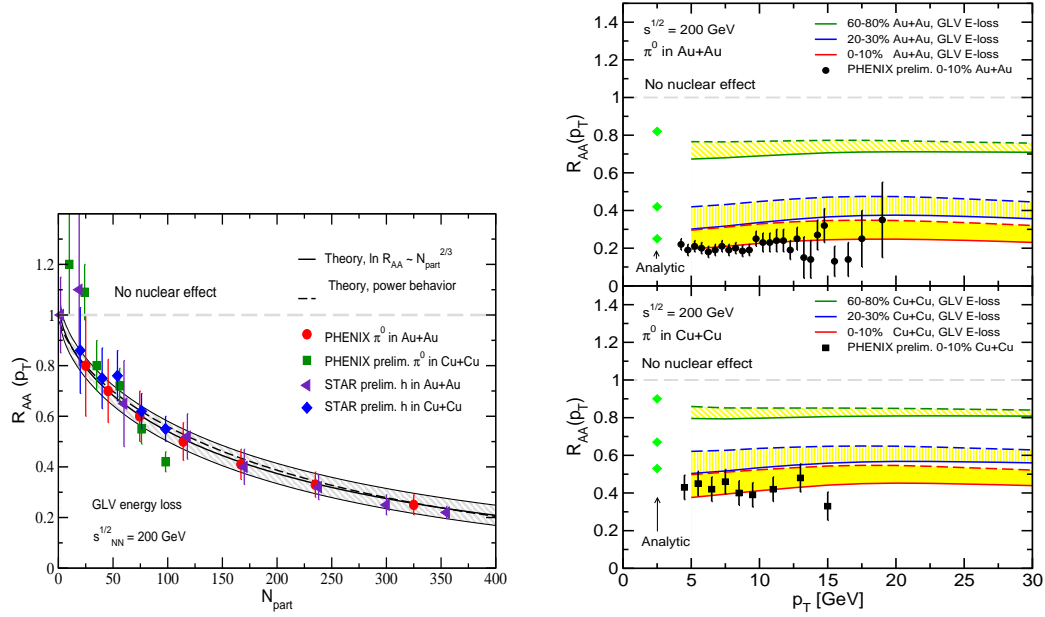

\includegraphics[width=2.6in,height=2.1in,angle=0]{Univ.eps}
\hspace*{0.5cm}
\includegraphics[width=2.5in,height=3.2in,angle=0]{CuAu.eps}
\caption{ Left panel:  comparison of the analytic $R_{AA}(p_T)$ to 
PHENIX and STAR data  on the $N_{\rm part}$-dependent quenching.
Right panel:  nuclear modification factor $R_{AA}$ for Au+Au
collisions at $\sqrt{s} = 200$~GeV versus $p_T$ and centrality.  
Similar calculation for Cu+Cu at $\sqrt{s} = 200$~GeV 
at RHIC. 
}
\label{analytic}
\end{figure}

Integrating over the gluon bremsstrahlung intensity  in the 
limit $2E/\mu^2 L \gg 1$ we find:   
\begin{eqnarray} \hspace*{-.5cm}
{ \Delta E }
&\approx&  
\frac{ C_R \alpha_s  }{4} 
\frac{\mu^2}{\lambda_g} L^2  \,  
\ln  \frac{2 E}{\mu^2 L}   \;,  \; \quad
{ \Delta E_T } 
\approx   C_R \frac{ 9 \pi \alpha_s^3  }{4} 
 \frac{1}{A_\perp} \frac{dN^{g}}{dy} \, L_T \,\,
 \ln   \frac{2 E_T}{\mu^2 L_T}   \; \qquad
\label{stat-de}
\end{eqnarray} 
for static and (1+1)D Bjorken-expanding plasmas,
respectively~\cite{Gyulassy:2000er,Gyulassy:2000fs}.
Tests of parton energy loss at RHIC and the 
LHC can be carried out via the centrality dependence
of jet quenching. With  ${dN^g}/{dy} \propto   N_{\rm part}$, 
$  L \propto N_{\rm part}^{2/3}$ and   $A_{\perp} 
\propto N_{\rm part}^{2/3}$, the GLV approach gives
a definite prediction for the system size dependence of
energy loss and jet quenching~\cite{Vitev:2006uc}: 
\begin{equation}
\epsilon = { \Delta E }/{E} \propto   
N_{\rm part}^{2/3} \;, \quad \ln R_{AA} \approx - k  N_{\rm part}^{2/3}  \; .
\label{RAA-log}
\end{equation}
Our analytic result for the  system size dependence of $R_{AA}(p_T)$
is shown in the left panel of Fig.~\ref{analytic}. It is fixed by the 
magnitude  of the suppression established in central Au+Au
collisions.  From this analysis
we expect a factor $\sim 2$ suppression in central Cu+Cu 
collisions.  Comparison to the  PHENIX
$p_T > 7$~GeV data~\cite{Shimomura:2005en}
and STAR  $p_T > 6$~GeV  data~\cite{Dunlop:2005xe} is also 
shown.

To verify the universal suppression pattern, Eq.~(\ref{RAA-log}), 
we carry out numerical simulations of $R_{AA}$ versus $p_T$ and 
centrality with QGP properties constrained by the entropy density, 
or equivalently $dN^g/dy$ and $\tau_0$ at RHIC. The 
modification of inclusive hadron production from final state
radiative energy loss can be represented as~\cite{Vitev:2006uc}:
\begin{eqnarray}
D_{h/c} (z) & \Rightarrow & \int_0^{1-z} d\epsilon \; P(\epsilon)  \; 
\frac{1}{1-\epsilon} D_{h/c} \left( \frac{z}{1-\epsilon} \right) 
+  \,  \int_z^{1} d\epsilon \; \frac{dN^g}{d \epsilon}(\epsilon) \;  
 \frac{1}{ \epsilon }  D_{h/g} \left( \frac{z}{\epsilon} \right) \;.
\qquad \label{nucmod-1}
\end{eqnarray}
In Eq.~(\ref{nucmod-1}) the first term is associated with the 
quenching of leading hadrons and  $P(\epsilon)$ is the 
probabilistic distribution for the fractional energy 
loss $\epsilon = \Delta E / E = \sum_{i = 1}^n \epsilon_i$, 
$\epsilon_i = \omega_i /E$,  due to multiple gluon 
emission. The right panel of Fig.~\ref{analytic}
shows the jet quenching results versus $p_T$ for three different
centralities  in Au+Au and Cu+Cu collisions at RHIC~\cite{Vitev:2006uc}.
Data is from PHENIX~\cite{Shimomura:2005en}.

\begin{figure}[!t]
\includegraphics[width=2.6in,height=2.1in,angle=0]{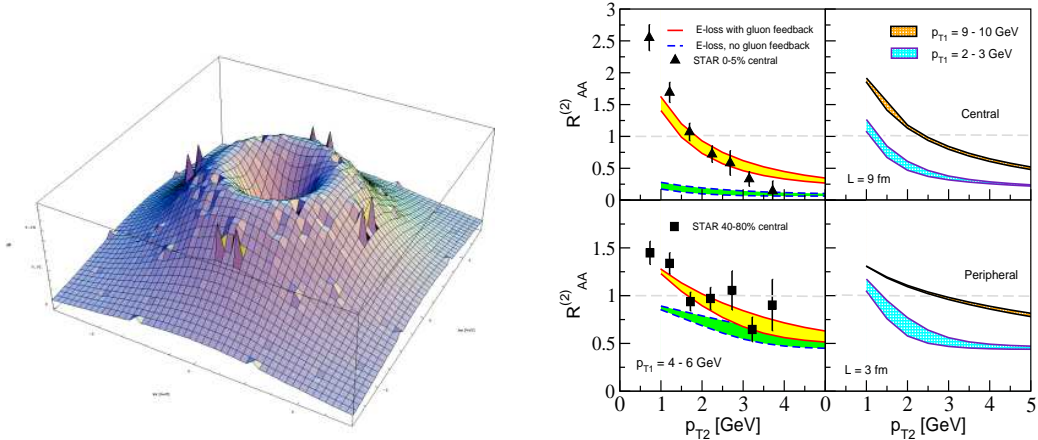}
\hspace*{0.5cm}
\includegraphics[width=2.5in,height=2.3in,angle=0]{feed.eps}
\caption{ Left panel: large angle distribution of the radiative 
gluons relative to th jet axis. Right panel:
 jet softening - redistribution of the 
energy from high $p_T$ to low $p_T$ particles~\cite{Vitev:2005yg}. 
Top and bottom panels illustrate central and 
peripheral collisions, respectively.
}
\label{feed}
\end{figure}

With fractional energy loss as large as $\sim 25\%$ for quarks and
$\sim 50\%$ for gluons~\cite{Vitev:2006uc}, the quenching of hard jets 
should be correlated with measurable enhanced production of low-$p_T$ 
hadrons~\cite{Adams:2005ph}. The second term in Eq.~(\ref{nucmod-1}) 
reflects the gluon feedback contribution, which for single 
inclusive measurements becomes important only at the LHC. For 
away-side di-hadron correlations this effect is already 
large at RHIC~\cite{Vitev:2005yg}. If follows from 
Eq.~(\ref{1storderSh}) that the angular behavior of the 
medium-induced gluon bremsstrahlung is significantly broader
when compared to  
vacuum radiation. Its most important feature is the cancellation 
for the collinear ${\bf k} \rightarrow {\bf q}_{\rm dir.}$ 
radiation. The left panel of Fig.~\ref{feed} shows the 
angular  distribution of $\omega=3$~GeV radiative gluon off 
a quark jet for $L=5$~fm, $\mu=1$~GeV, 
$\lambda_g = 1$~fm~\cite{Vitev:2005yg}. 
The remnant of the leading parton is not shown.

Depletion of hadrons from the quenching of the parent parton 
alone  leads to a large suppression of the 
double inclusive cross section with weak $p_{T_2}$ dependence,
shown in the right panel of Fig.~\ref{feed}.
Hadronic feedback from the  
medium-induced gluon radiation, however, completely changes 
the nuclear di-hadron modification factor $R^{(2)}_{AA}$.     
It now shows a clear transition from a quenching of 
the away-side jet at high transverse momenta to enhancement
at  $p_{T_2} \leq  2$~GeV~\cite{Vitev:2005yg} compatible with 
STAR measurements~\cite{Adams:2005ph}.

\begin{figure}[!t]
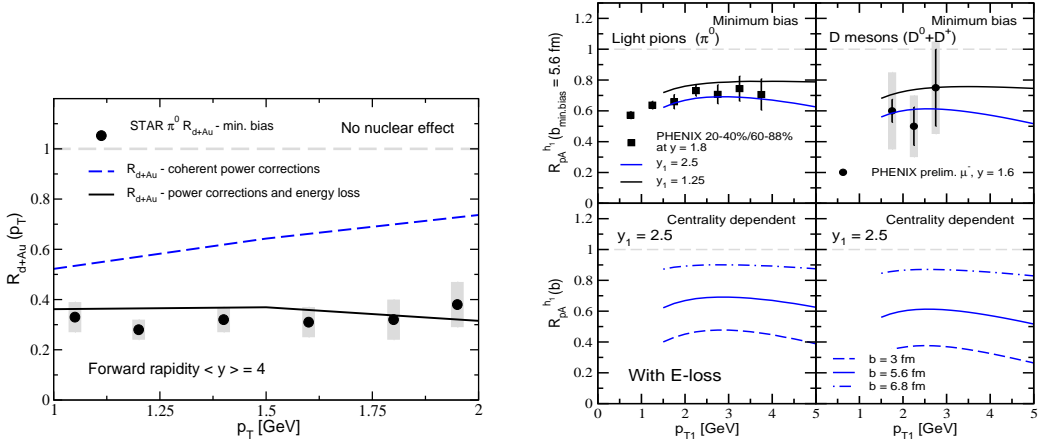

\includegraphics[width=2.5in,height=1.8in,angle=0]{STAR-y4.eps}
\hspace*{0.5cm}
\includegraphics[width=2.6in,height=2.3in,angle=0]{Power-Charm-EL-NEW.eps}
\caption{ Left panel: calculated suppression of $\pi^0$ production 
at $y=4 $ for 
d+Au collisions at $\sqrt{s_{NN}}=200$~GeV at  RHIC. 
Right panel: nuclear modification of single inclusive $\pi^0$ 
(left) and $D^0 + D^+$  mesons (right) in d+Au collisions at RHIC 
versus rapidity and centrality. 
}
\label{rapi}
\end{figure}

The physics process  that has so far not been addressed is 
initial state  energy loss in cold nuclear 
matter~\cite{Vitev:2006bi,Vitev:prep}. To first order in opacity: 
\begin{eqnarray}
\omega\frac{dN_g}{d\omega\, d^2 {\bf k}}  &=&
\frac{ C_R \alpha_s}{  \pi^2 } 
\int_{0}^L \frac{d \Delta z}{\lambda_g(z)}  
\int  d^2 {\bf q} \, 
\frac{\mu^2(z)}{\pi({\bf q}^2 + \mu^2(z) )^2} \times  \nonumber \\ 
&& \hspace*{+1cm} \times 
 \left[ \frac{{\bf q}^2}{{\bf k}^2 ({\bf k}-{\bf q})^2} 
-  2    \frac{{\bf q}^2 - {\bf k} \cdot {\bf q} }
{{\bf k}^2 ({\bf k}-{\bf q})^2}    
\cos \left( \frac{  {\bf k}^2  \Delta z}{2\omega} \right)  \right] 
\;.
\label{1stordIS}
\end{eqnarray}
From Eq.~(\ref{1stordIS}) with the requirement 
$\omega {dN_g}/{d\omega\, d^2 {\bf k}} \geq 0$ we can 
evaluate the energy loss and  implement it in the 
PQCD hadron production formalism:
\begin{eqnarray}
&&  \epsilon = \frac{\Delta E}{E} \propto \frac{L}{\lambda} 
= \kappa A^{1/3}  \;,  \qquad   \phi_{q,g/N}(x,Q^2) \rightarrow 
\phi_{q,g/N}\left( \frac{x}{1-\epsilon}, Q^2 \right) \;.
\label{i-eloss}
\label{implIS}
\end{eqnarray} 
As shown in the  left panel of Fig.~\ref{rapi}, 
shadowing calculations can only account for $\sim 50\%$ 
of the hadron suppression at forward rapidities p+A 
reactions at RHIC~\cite{Vitev:2006bi,Qiu:2004da}. 
Taking into account initial-state energy loss, 
we find good description of the $y=4$ $R_{AA}(p_T)$ 
for $\pi^0$s by STAR~\cite{Adams:2006uz}. The right panel 
of  Fig.~\ref{rapi} shows the same perturbative calculation
for light hadrons and heavy mesons~\cite{Vitev:2006bi} 
at intermediate rapidities 
$y \sim 2$ and  $\sqrt{s}=200$~GeV d+Au collisions at 
RHIC compared to PHENIX data~\cite{Adler:2004eh}.

\section{Conclusions}

Dynamical calculations of initial- and final-state 
elastic~\cite{Vitev:2003xu}, 
inelastic~\cite{Gyulassy:2000er,Gyulassy:2000fs}
and coherent~\cite{Qiu:2003vd,Qiu:2004da} multiple 
parton scattering have been incorporated 
in the framework of the perturbative QCD factorization 
approach to describe high $p_T$ and $E_T$ observables in
collisions of heavy nuclei. They naturally account for 
the process  dependence of nuclear effects~\cite{Vitev:2006bi} 
and provide adequate unified description of seemingly disparate 
phenomena, such as the Cronin effect~\cite{Accardi:2002ik} and 
nuclear shadowing~\cite{Armesto:2006ph} in p+A / $\ell$ + A 
reactions and high-$p_T$ jet 
quenching~\cite{Gyulassy:2003mc,Vitev:2006uc} and 
low $p_T$ large-angle di-hadron enhancement~\cite{Vitev:2005yg}
in A+A reactions. In cold nuclear matter, parton diffusion, 
nuclear enhanced power corrections and initial state energy 
loss  are  compatible with similar momentum transfer scales: 
$\xi^2 A^{1/3} \approx 2 \mu^2  L / \lambda $. In the QGP, jet 
quenching results are in agreement  with the predicted 
energy and centrality $\propto N_{\rm part}^{2/3}$ dependence 
of the non-Abelian parton energy loss.

{\bf Acknowledgments:}
This research is supported in part by the US Department 
of Energy under Contract No. W-7405-ENG-3
and the J. Robert Oppenheimer Fellowship of the 
Los Alamos National Laboratory.

\end{document}